\documentclass[aps,prl,twocolumn,superscriptaddress,amsmath,amssymb,preprintnumbers,showpacs]{revtex4}

\usepackage{graphicx}
\usepackage{dcolumn}

\begin{document}


\title{Quantum theory of heating of a single trapped ion}


\author{F. Intravaia}

\affiliation{INFM and MIUR, Dipartimento di Scienze Fisiche ed
Astronomiche dell'Universit\`{a} di Palermo, via Archirafi 36,
90123 Palermo, Italy.}

\author{S. Maniscalco}
\email{sabrina@fisica.unipa.it}

\affiliation{INFM and MIUR, Dipartimento di Scienze Fisiche ed
Astronomiche dell'Universit\`{a} di Palermo, via Archirafi 36,
90123 Palermo, Italy.}

\author{J. Piilo}

\affiliation{Helsinki Institute of Physics, PL 64, FIN-00014
Helsingin yliopisto, Finland}

\author{A. Messina}

\affiliation{INFM and MIUR, Dipartimento di Scienze Fisiche ed
Astronomiche dell'Universit\`{a} di Palermo, via Archirafi 36,
90123 Palermo, Italy.}

\date{\today}

\begin{abstract}
The heating of trapped ions due to the interaction with a {\it
quantized environment} is studied { \it without performing the
Born-Markov approximation}. A generalized master equation local in
time is derived and a novel theoretical approach to solve it
analytically is proposed. Our master equation is in the Lindblad
form with time dependent coefficients, thus allowing the
simulation of the dynamics by means of the Monte Carlo Wave
Function (MCWF) method.
\end{abstract}


\maketitle

 Single ions confined in miniaturized radio frequency traps can be
cooled down by means of laser cooling techniques very efficiently
\cite{cooling}. Experiments have shown that the center of mass of
a single cooled trapped ion undergoes a quantized harmonic
oscillatory motion \cite{qho}. During the
 last decade a great deal of attention has been devoted to such
 systems since they turn out to be very weakly coupled to external
 environment, thus allowing the manipulation and observation of the
 coherent dynamics of a single quantum system \cite{manipulation}. Indeed, by
 irradiating the ion with properly configured laser beams, it is
 possible to manipulate the vibronic state of the ion. The
 easiness to engineer at will the interaction between the motional
 and internal degrees of freedom of the ion makes it possible to
 perform experimental tests of fundamental features of quantum
 mechanics. For example, many nonclassical states of the
 oscillatory motion of the ion have been prepared and measured \cite{nonclass}. By
 using multiple ions in a linear Paul trap, experiments on quantum
 nonlocality have been performed \cite{nonlocality} and many-particle entangled
 states have been realized \cite{entangl}. Furthermore cold trapped ions have
 been recently proposed as a physical implementation for quantum
 computation \cite{qcompu}.

 Both for fundamental studies and for technological
 applications it is of great interest to understand and study
 those factors limiting the fidelity of the operations performed to manipulate coherently the quantum state of the ion.
Heating of the center of mass oscillatory motion of one or more
trapped ions seems to be one of the major practical sources of
decoherence \cite{heating}. The process of heating of a single
trapped ion is due to the electromagnetic coupling between noisy
electric fields and the ion. Such fields give rise to fluctuating
forces acting on the ion which can cause an increase in its
motional energy. The physical origin of the noisy electric fields
in the center of the trap has not yet been unambiguously
identified. In fact, measurements of the heating rate are very
difficult to perform since they require high sensitivity and they
may depend on parameters related to the specific trap geometry
used \cite{heating}. For this reason a general theory of heating
would help in identifying and possibly reduce the up to now
unknown sources of noise.

Most of the theoretical studies on the heating of trapped ions
deal with the interaction between a single harmonic oscillator and
a reservoir which is either described as a classical stochastic
field
\cite{heatingtheory1,heatingtheory2,heatingtheory3,budimatos} or
as an infinite chain of quantum harmonic oscillators. In the first
case an analytic solution for the dynamics of the system can be
found without performing the Born-Markov approximation
\cite{heatingtheory1,heatingtheory3}. In other words, it is
possible to study the short time evolution of the heating function
when the field noise spectrum is not flat, as it is indeed in any
real experimental situation. However the analytical treatments, in
this case, cannot describe the thermalization process,  i.e. the
long time behavior, since the stochastic field continuously feeds
energy in the system, leading to an infinite growth of the heating
function. On the other hand, modelling a thermal reservoir as an
infinite chain of harmonic oscillators at  temperature $T$ allows,
under certain approximations, to describe the dynamics of the
system in terms of a master equation of the form
\cite{heatingmarkov}
\begin{eqnarray}
\frac{ \partial \hat{\rho}}{\partial t}= &-& \Gamma (N+1) \left[
\hat{a}^{\dag} \hat{a} \hat{\rho} - 2 \hat{a} \hat{\rho}
\hat{a}^{\dag} + \hat{\rho} \hat{a}^{\dag} \hat{a} \right]
\nonumber \\
&-& \Gamma N \left[  \hat{a} \hat{a}^{\dag} \hat{\rho} - 2
\hat{a}^{\dag} \hat{\rho} \hat{a} + \hat{\rho} \hat{a}
\hat{a}^{\dag}
 \right], \label{eq:1}
\end{eqnarray}
where $\hat{a}^{\dag}$ and $\hat{a}$ are the creation and
annihilation operators of vibrational quanta, $\Gamma$ is the
heating constant and the environment is assumed to be at thermal
equilibrium, $N$ being the mean number of reservoir excitations at
 temperature $T$. In Eq. (\ref{eq:1}) the terms proportional to $
\Gamma (N+1)$ account for spontaneous and stimulated emission,
that is a transfer of energy from the system to the reservoir,
while the terms proportional to $\Gamma N$ account  for
absorption, that is a transfer of energy from the reservoir to the
system. The presence of the spontaneous emission term, arising
from a quantized description of the reservoir, ensures the
thermalization process. Indeed, from equation (\ref{eq:1}), it is
possible to derive the following expression describing the time
evolution of the heating function:
\begin{equation}
\langle \hat{n} (t) \rangle = n(\omega_0) \left( 1- e^{- \Gamma
t}\right), \label{eq:1a}
\end{equation}
with $n(\omega_0)= \left( e^{\beta \hbar \omega_0
}-1\right)^{-1}$, $\beta=(KT)^{-1}$ and $\omega_0$ frequency of
the trap. In order to describe the dynamics of the system by means
of the master equation (\ref{eq:1}), however, it is necessary to
assume the validity of the Born-Markov approximation. As a
consequence such a treatment is valid only for times $t$ much
bigger than the correlation time of the reservoir and thus does
not give the correct short time behavior.

The aim of our paper is to study the heating of a single trapped
ion interacting with a quantized reservoir {\it without performing
the Born-Markov approximation}. One of the main results of the
paper is the derivation of an analytical expression for the
heating function describing both the non-Markovian short time
behavior and the asymptotic thermalization process.

The Hamiltonian describing the interaction between the ion motion
and a quantized reservoir modeled as an infinite chain of harmonic
oscillators, can be written as follows $ \hat{H}= \hbar
\sqrt{\alpha} \sum_{j=0}^{\infty} c_j \big( \hat{a} +
\hat{a}^{\dag} \big) \big( \hat{b}_j + \hat{b}_j^{\dag} \big) $,
where $\hat{b}_j$ is the annihilation operator of reservoir
excitations and $\alpha$ is the coupling strength. By using the
time-convolutionless projection operator technique \cite{timeconv}
and performing the Rotating Wave Approximation, we have derived
the following master equation describing the non-Markovian
dynamics of the system in the interaction picture:
\begin{eqnarray}
\frac{ \partial \hat{\rho}}{\partial t}= &-& \frac{\bar{\Delta}(t)
+ \gamma (t)}{2} \left[ \hat{a}^{\dag} \hat{a} \hat{\rho} - 2
\hat{a} \hat{\rho} \hat{a}^{\dag} + \hat{\rho} \hat{a}^{\dag}
\hat{a} \right]
\nonumber \\
&-& \frac{\bar{\Delta}(t) - \gamma (t)}{2} \left[  \hat{a}
\hat{a}^{\dag} \hat{\rho} - 2 \hat{a}^{\dag} \hat{\rho} \hat{a} +
\hat{\rho} \hat{a} \hat{a}^{\dag}
 \right]. \label{eq:3}
\end{eqnarray}
The functions $\bar{\Delta}(t)$ and $\gamma (t)$ are defined in
terms of the two reservoir characteristic functions
\cite{cfunction}: the correlation function $\kappa (\tau) = \alpha
\int_0^{\infty} \omega |g(\omega)|^2 (2 n(\omega)+1) \cos (\omega
\tau) d \omega$ and the susceptibility $\mu (\tau)= \alpha
\int_0^{\infty} \omega |g(\omega)|^2 \sin (\omega \tau) d \omega$,
where $n(\omega)=\left(e^{\beta \hbar \omega}-1\right)^{-1}$.
Moreover, assuming a Lorentzian spectral density $|g(\omega)|^2 =
\omega_c^2 / [\pi (\omega^2 + \omega_c^2)]$ (Ohmic environment
\cite{weiss}) with $\omega_c$ reservoir frequency cut, in the high
temperature regime defined by the condition $(\hbar \beta )^{-1}
\gg \omega_c $ and in the weak coupling limit, the real functions
$\bar{\Delta} (t) $ and $\gamma(t)$ can be written in a simple
analytic form:
\begin{eqnarray}
\bar{\Delta} (t) \!\!\!&=& \! \!\! \frac{\alpha}{\hbar \beta
\omega_0} \frac{r}{r^2+1}\! \Big\{ e^{- \omega_c t}\! \sin(
\omega_c t / r) \! \nonumber \\
&+& \! r \! \left[ 1\!-\!  e^{- \omega_c t} \! \cos(\omega_c t
/r ) \!\right] \!\Big\}, \label{eq:4}\\
\gamma (t) \!\!\! &=& \!\!\! \frac{\alpha}{2} \frac{r^2}{r^2+1}
\Big[ \! 1 \!- \! e^{- \omega_c t} \cos(\omega_c t /r) \nonumber
\\ &-& r e^{- \omega_c t}\! \sin( \omega_c t /r) \! \Big],
\label{eq:5}
\end{eqnarray}
where  $r= \omega_c / \omega_0$. Remembering that in the high
temperature regime $n(\omega_0)\simeq (\hbar \beta \omega_0)^{-1}
$, it is easy to check that in the Markovian limit, that is when
$t \gg \tau_R $ with $\tau_R= 1/ \omega_c$ correlation time of the
reservoir, equation (\ref{eq:3}) reduces to equation (\ref{eq:1}).
The analytic expression of $\bar{\Delta} (t) $ and $\gamma(t)$ for
a generic temperature $T$ is, in general, more complicated.
Nonetheless the temperature dependence of these coefficients of
the master equation brings to light interesting features in the
dynamics of the system and, for this reason, it is carefully
discussed in a follow up paper \cite{inprep}.

Our non-Markovian master equation (\ref{eq:3}) has the general
form of the time-convolutionless master equations
\cite{petruccione,petruccionebook}. However, it is worth noting
that it is not only local in time but also, as long as the
coefficients $\bar{\Delta}(t) + \gamma (t)$ and $\bar{\Delta}(t) -
\gamma (t)$ are positive,  in the Lindblad form. We have verified
that, for the value of parameters currently used in the
experiments and for $r >1$ the time dependent sum and difference
coefficients are positive at all times $t$. Note that the
condition $r > 1$ simply means that the spectral density of the
reservoir overlaps with the frequency of oscillation of the ion,
which is a reasonable and commonly done assumption. In other
words, we demonstrate for the first time that the non-Markovian
dynamics of the reduced density matrix of the system here
considered, is described by a Lindblad-type master equation. Thus,
it is possible to use the standard MCWF method to unravel Eq.
(\ref{eq:3}) \cite{MCWF}. We note that up to now, only quantum
state diffusion unravelings have been known and used for
simulating the temporal behavior of a harmonic oscillator
interacting with a non-Markovian quantized environment \cite{QSD}.
We would like to emphasize that the advantage of working with a
master equation of Lindblad type whose analytical solution may be
successfully found, as in our case, is that the positivity  of the
density matrix is  ensured. In the following we will use the
quantum jump MCWF method to obtain the temporal behavior of the
heating function and compare it with the analytic solution.

We now briefly sketch the novel analytical treatment we have
developed to solve the master equation (\ref{eq:3}) and derive a
closed expression for the heating function describing both its
short and its long time behavior. A detailed description of the
method will be given elsewhere \cite{inprep}. The first step of
our approach consists in expanding the density matrix of the
system as follows \cite{vogbook}:
\begin{equation}
\hat{\rho}(t)= \frac{1}{\pi} \int \chi_t (\lambda, \lambda^*)
\exp\left( \lambda^* \hat{a} - \lambda \hat{a}^{\dag} \right) d^2
\lambda \label{eq:6}
\end{equation}
In Eq. (\ref{eq:6}), $\lambda \in \mathbb{C}$ and $\chi_t
(\lambda, \lambda^*)= Tr \left\{\exp  \left( \lambda
\hat{a}^{\dag}  - \lambda^* \hat{a} \right) \hat{\rho}(t)
\right\}$ is the quantum characteristic function. Inserting Eq.
(\ref{eq:6}) into Eq. (\ref{eq:3}) and by using some useful
algebric properties of the cronologically ordered time evolution
superoperators ${\bf T}_c(t)$ of the system, defined by
$\hat{\rho}(t) = {\bf T}_c(t) \hat{\rho}(0) $, we obtain the
following expression for the quantum characteristic function:
\begin{equation}
\chi_t (\lambda, \lambda^*) \!=\! \exp \left( \! - \Delta(t)
|\lambda|^2 \right)\! \chi_0 \! \left( T_{\gamma}^{1/2}(t)
\lambda; T_{\gamma}^{1/2}(t) \lambda^* \!\right)\!\!. \label{eq:7}
\end{equation}
\begin{figure}
\includegraphics{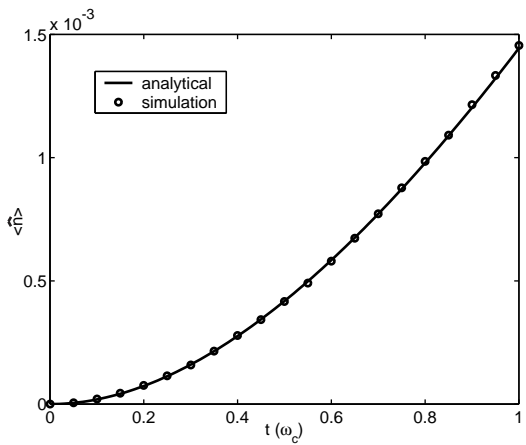}
\caption{\label{fig:1} Short time behavior of the heating function
 for coupling constant  $\alpha= 0.1 Hz$,
 trap frequency $\omega_0= 10^7 Hz$, {\it r}=10 and $T=300 K$. We
compare the analytic solution and the MCWF simulation with $10^7$
histories.}
\end{figure}
\begin{figure}
\includegraphics{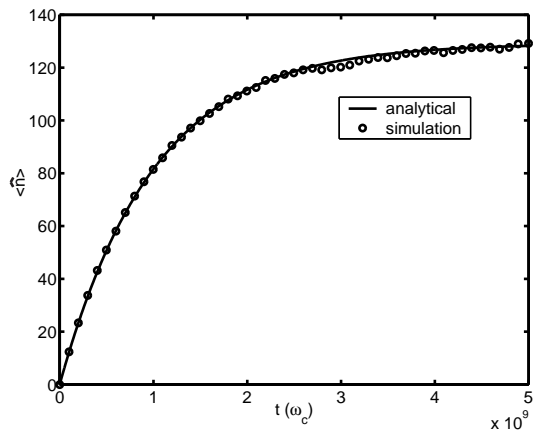}
\caption{\label{fig:2} Long time behavior of the heating function
 showing the thermalization process, for
coupling constant $\alpha= 0.1 Hz$,  trap frequency $\omega_0=
10^7 Hz$, {\it r}=10 and $T=10$ mK. We compare the analytic
solution and the MCWF simulation with $10^4$ histories. We note
that, even for such a small value of the reservoir temperature,
the high temperature approximation holds since $ (\hbar
\beta)^{-1} \simeq 10^9 Hz$ which is one order of magnitude bigger
than $\omega_c$. }
\end{figure}
In this equation $\chi_0$ is the quantum characteristic function
at the initial time instant $t=0$,
\begin{equation}
T_{\gamma} (t) = \exp \left( - 2 \int_0^{t} \gamma(t') dt'
\right), \label{eq:71bis}
\end{equation}
and
\begin{equation}
\Delta(t) = \int_0^{t} T_{\gamma}(t) T_{\gamma}^{-1}(t')
\bar{\Delta}(t'). \label{eq:7bis}
\end{equation}
The functions $\bar{\Delta} (t)$ and $\gamma (t) $ are defined by
eqs. (\ref{eq:4}) and (\ref{eq:5}). For example, assuming as
initial state a vibrational Fock state $\vert k \rangle$, Eq.
(\ref{eq:7}) becomes
\begin{equation}
\chi_t (\lambda, \lambda^*) \!=\! \exp \!\left[ \!-\! \left( \!
\Delta(t) \!+\! \frac{T_{\gamma} (t)}{2} \right) |\lambda|^2
\right]\! \! L_k \left(T_{\gamma} (t) |\lambda|^2 \!\right)\!\!,
\label{eq:8}
\end{equation}
where $L_k \left( y \right)$ is the Laguerre polynomial of order
$k$ in $y$. Eq. (\ref{eq:7}) together with Eq. (\ref{eq:6}) gives
the {\it analytic expression of the density matrix of the system
in an operatorial form at any time instant $t$ }. From Eq.
(\ref{eq:7}) it is easy to derive the following expression for the
mean vibrational quantum number, i.e. for the heating function:
\begin{equation}
\langle \hat{n} (t) \rangle = \Delta (t) + \frac{1}{2} \left(
T_{\gamma} (t) -1 \right) + k T_{\gamma} (t). \label{eq:10}
\end{equation}
In Figs.~\ref{fig:1} and \ref{fig:2} we show the short and long
time behavior of the heating function of a trapped ion initially
prepared in its vibrational ground state. We compare the
analytical solution and the MCWF simulation performed starting
from the master equation (\ref{eq:3}). The figures show a very
good agreement between the analytical and the numerical
approaches. We note that our quantum theory of heating predicts
the initial quadratic behavior typical of non-Markovian dynamics.
In Fig.~{\ref{fig:3} we compare the non-Markovian and Markovian
time evolution for times $t$ smaller than the reservoir
correlation time $\tau_R = 1/\omega_c$. A similar result was
deduced by James and by Budini in the case of a classical
environment described in terms of a stochastic noisy electric
field \cite{heatingtheory1,heatingtheory3}. Our approach, however,
allows also to describe the asymptotic thermalization process as
shown in Fig.~{\ref{fig:2}. Indeed, for times $t \gg \tau_R$ and
for a flat reservoir spectrum $r\gg 1$, the heating function given
by equation (\ref{eq:10}) reduces to the well known Markovian
equation given by Eq. (\ref{eq:1a}), as one can easily derive
substituting the asymptotic expressions for
$\bar{\Delta}(t\rightarrow \infty)$ and $\gamma(t\rightarrow
\infty)$ into equations (\ref{eq:71bis}), (\ref{eq:7bis}) and
(\ref{eq:10}), in the limit $r\gg 1$.
\begin{figure}
\includegraphics{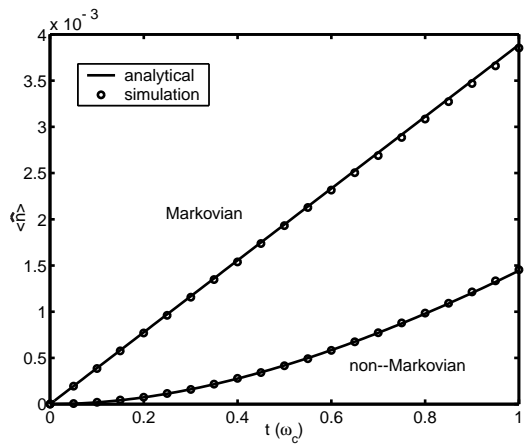}
\caption{\label{fig:3} Comparison between the short time Markovian
and non-Markovian time evolutions of the heating function for
coupling constant  $\alpha= 0.1 Hz$,  trap frequency $\omega_0=
10^7 Hz$, {\it r}=10 and $T=300 K$.}
\end{figure}

In conclusion we have proposed a new approach for studying the
heating of a single trapped ion due to the interaction with a
quantized reservoir. Our approach does not rely on the Born-Markov
approximation and thus allows to describe the initial quadratic
behavior of the heating function. Moreover, due to the quantum
description of the environment, our solution describes correctly
the thermalization process. Stated another way our quantum theory
of heating bridges a gap existing in the literature. Until now,
indeed, only non-Markovian analytical solutions, in the case of
interaction with a classical environment, or Markovian analytical
solutions, for interactions with a quantum environment, were
known.

It is important to emphasize that measurements of the initial
quadratic behavior of the heating function are difficult to
perform for two reasons. Firstly because this would mean to make
many measurements in an interval of time of the order of $\tau_R
\simeq 10^{-8}$ s, for the value of parameters used in the paper.
The second and more important reason is related to the difficulty
in revealing so small variations in $\langle \hat{n} (t) \rangle$
as the ones shown in Fig.~{\ref{fig:1}. Very recently, however, it
has been experimentally demonstrated the possibility of
engineering both the type of reservoir interacting with a single
trapped ion and the coupling between the system and the
environment \cite{engineeringres}. The analytical solution we have
presented in this paper makes it possible to look for ranges of
the relevant parameters of both the system and the reservoir in
correspondence of which the non-Markovian quadratic behavior
becomes experimentally observable.

The experimental ability in engineering reservoir suggests also
another application. As we have already mentioned in the paper, we
have analyzed the dependence of the  coefficients of our master
equation (\ref{eq:3}) on reservoir parameters such as its
temperature $T$ and frequency cut $\omega_c$. We have found that
by changing such parameters the master equation passes from
Lindblad type to non-Lindblad type. Such a modification does
reflect a deep change in the dynamics of the system. This result
will be discussed in more  detail in  a follow up paper since we
believe that the possibility of studying analytically such changes
may give more insight in understanding the fundamental properties
of the heating process of single trapped ions.

\acknowledgments We acknowledge K.-A. Suominen for helpful
comments and the Finnish Center for Scientific Computing (CSC) for
computing resources. S.M. acknowledges the Helsinki Institute of
Physics, where part of the work was done, for the hospitality.
J.P. acknowledges financial support from the National Graduate
School on Modern Optics and Photonics.

\end{document}